\documentclass[doublecol]{epl2} 
\usepackage{verbatim}
\usepackage{graphicx}
\usepackage[overload]{textcase}
\DeclareGraphicsExtensions{.eps,.pdf,.png,.gif,.jpg,.mht}
\usepackage{multirow}
\usepackage{stfloats}
\usepackage{hhline}
\usepackage{arydshln}
\usepackage{amssymb,amsmath}
\usepackage{verbatim}
\usepackage{url}
\RequirePackage[htt]{hyphenat}      % Sillabazione per carattere monospazio 

\title{Measurement of the forward charged particle pseudorapidity density in $pp$ collisions at $\sqrt{s} = 7$ TeV with the TOTEM experiment}
\shorttitle{Measurement of the forward charged particle pseudorapidity density in $pp$ collisions at $\sqrt{s} = 7$ TeV}

\author{
The TOTEM Collaboration\\
G.~Antchev\thanks{INRNE-BAS, Institute for Nuclear Research and Nuclear Energy, Bulgarian Academy of Sciences, Sofia, Bulgaria.}\addtocounter{footnote}{-1}
\and P.~Aspell\inst{8}
\and I.~Atanassov\inst{8}\hspace{-0.15cm}\footnotemark
\and V.~Avati\inst{8}
\and J.~Baechler\inst{8}
\and V.~Berardi\inst{5b,5a}
\and M.~Berretti\inst{7b}
\and E.~Bossini\inst{7b}
\and M.~Bozzo\inst{6b,6a}
\and P.~Brogi\inst{7b}
\and E.~Br\"{u}cken\inst{3a,3b}
\and A.~Buzzo\inst{6a}
\and F.~S.~Cafagna\inst{5a}
\and M.~Calicchio\inst{5b,5a}
\and M.~G.~Catanesi\inst{5a}
\and C.~Covault\inst{9}
\and M.~Csan\'{a}d\inst{4}
\and T.~Cs\"{o}rg\H{o}\inst{4}
\and M.~Deile\inst{8}
 \and K.~Eggert\inst{9}
 \and V.~Eremin\thanks{Ioffe Physical - Technical Institute of Russian Academy of Sciences.}
 \and R.~Ferretti\inst{6a,6b}
 \and F.~Ferro\inst{6a}
 \and A. Fiergolski\thanks{Warsaw University of Technology, Poland.}
 \and F.~Garcia\inst{3a}
 \and S.~Giani\inst{8}
 \and V.~Greco\inst{7b,8}
 \and L.~Grzanka\inst{8}\hspace{-0.15cm}\thanks{Institute of Nuclear Physics, Polish Academy of Science, Cracow, Poland.}\addtocounter{footnote}{-2}
 \and J.~Heino\inst{3a}
 \and T.~Hilden\inst{3a,3b}
 \and M.~R.~Intonti\inst{5a}
 \and J.~Ka\v{s}par\inst{1a,8}
 \and J.~Kopal\inst{1a,8}
 \and V.~Kundr\'{a}t\inst{1a}
 \and K.~Kurvinen\inst{3a}
 \and S.~Lami\inst{7a}
 \and G.~Latino\inst{7b}
 \and R.~Lauhakangas\inst{3a}
 \and  T.~Leszko\footnotemark
 \and E.~Lippmaa\inst{2}
 \and M.~Lokaj\'{\i}\v{c}ek\inst{1a}
 \and M.~Lo~Vetere\inst{6b,6a}
 \and F.~Lucas~Rodr\'{i}guez\inst{8}
 \and M.~Macr\'{\i}\inst{6a}
 \and L.~Magaletti\inst{5b,5a}
 \and T.~M\"aki\inst{3a}
 \and A.~Mercadante\inst{5b,5a}
 \and N.~Minafra\inst{8} 
 \and S.~Minutoli\inst{6a}\addtocounter{footnote}{1}
 \and F.~Nemes\inst{4}\hspace{-0.15cm}\thanks{Department of Atomic Physics, ELTE University, Hungary.}
 \and H.~Niewiadomski\inst{8}
 \and E.~Oliveri\inst{7b}
 \and F.~Oljemark\inst{3a,3b}
 \and R.~Orava\inst{3a,3b}
 \and M.~Oriunno\inst{8}\hspace{-0.15cm}\thanks{SLAC National Accelerator Laboratory, Stanford CA, USA.}
 \and K.~\"{O}sterberg\inst{3a,3b}
 \and P.~Palazzi\inst{7b}
 \and J.~Proch\'{a}zka\inst{1a}
 \and M.~Quinto\inst{5a}
 \and E.~Radermacher\inst{8}
 \and E.~Radicioni\inst{5a}
 \and F.~Ravotti\inst{8}
 \and E.~Robutti\inst{6a}
 \and L.~Ropelewski\inst{8}
 \and G.~Ruggiero\inst{8}
 \and H.~Saarikko\inst{3a,3b}
 \and A.~Santroni\inst{6b,6a}
 \and A.~Scribano\inst{7b}
 \and W.~Snoeys\inst{8}
 \and J.~Sziklai\inst{4}
 \and C.~Taylor\inst{9}
 \and N.~Turini\inst{7b}
 \and V.~Vacek\inst{1b}
 \and M.~Vitek\inst{1b}
 \and J.~Welti\inst{3a,3b}
 \and J.~Whitmore\inst{10}
 }          %ends author list
\shortauthor{The TOTEM Collaboration (G.~Antchev \etal)}
\institute{
\inst{1a} {Institute of Physics of the Academy of Sciences of the Czech Republic, Praha, Czech Republic.}\\
\inst{1b} {Czech Technical University, Praha, Czech Republic.}\\
\inst{2} {National Institute of Chemical Physics and Biophysics NICPB, Tallinn, Estonia.}\\
\inst{3a}{Helsinki Institute of Physics, Finland.}\\
\inst{3b}{Department of Physics, University of Helsinki, Finland.}\\
\inst{4} {MTA Wigner Research Center, RMKI, Budapest, Hungary.}\\
\inst{5a}{INFN Sezione di Bari, Italy.}\\
\inst{5b}{Dipartimento Interateneo di Fisica di Bari, Italy.}\\
\inst{6a}{Sezione INFN, Genova, Italy.}\\
\inst{6b}{Universit\`{a} degli Studi di Genova, Italy.}\\
\inst{7a}{INFN Sezione di Pisa, Italy.}\\
\inst{7b}{Universit\`{a} degli Studi di Siena and Gruppo Collegato INFN di Siena, Italy.}\\
\inst{8} {CERN, Geneva, Switzerland.}\\
\inst{9} {Case Western Reserve University, Dept. of Physics, Cleveland, OH, USA.}\\
\inst{10}{Penn State University, Dept. of Physics, University Park, PA, USA.}\\
}             %ends institute list

\abstract{
The TOTEM experiment has measured the charged particle pseudorapidity density dN$_{\textnormal{ch}}$/d$\eta$ in $pp$ collisions at $\sqrt{s} =$ 7 TeV for $5.3<|\eta|<6.4$ in events with at least one charged particle with transverse momentum above 40 MeV/c in this pseudorapidity range. This extends the analogous measurement performed by the other LHC experiments to the previously unexplored forward $\eta$ region. The measurement refers to more than  99\% of non-diffractive processes and to single and double diffractive processes with diffractive masses above $\sim\,$3.4 GeV/$c^{2}$, corresponding to about 95\% of the total inelastic cross-section. The dN$_{\textnormal{ch}}$/d$\eta$ has been found to decrease with $|\eta|$, from 3.84 $\pm$ 0.01(stat) $\pm$ 0.37(syst) at $|\eta| = 5.375$  to 2.38 $\pm$ 0.01(stat) $\pm$ 0.21(syst) at $|\eta| = 6.375$. Several MC generators have been compared to data; none of them has been found to fully describe the measurement.
}

\begin{document}
\maketitle

\section{Introduction}
The pseudorapidity density of charged particles produced in high energy proton-proton ($pp$) collisions reflects the strong interaction dynamics that is only partly described by perturbative QCD. Non-perturbative models and parametrisations are used in the Monte Carlo (MC) event generators to describe the hadronisation of the partonic final states and to model diffractive processes\cite{Skand,RistoRef}. In the forward region, where peripheral diffractive processes are important, the uncertainties are pronounced. A better understanding of these effects is also important for the interpretation of high energy showers recorded by cosmic ray experiments\cite{eggert,CommonTDR,Denterria}.
A direct measurement of the forward particle densities is, therefore, extremely valuable in constraining the theoretical models for particle production in $pp$ interactions.
The measurement of the charged particle pseudorapidity density (dN$_{\textnormal{ch}}$/d$\eta$) in the range $5.3<|\eta|<6.4$ is presented here. This quantity is defined as the mean number of charged particles per single $pp$ collision and unit of pseudorapidity $\eta$, where $\eta\,\equiv\,-\textnormal{ln}[\textnormal{tan}(\theta/2)]$, and $\theta$ is the polar angle of the direction of the particle with respect to the counterclockwise beam direction.

\section{Experimental apparatus}
TOTEM is a dedicated experiment to measure the total cross section, elastic scattering and diffractive processes at the LHC\cite{LETTEROfINTENT,TOTEM_TDR}. The experimental apparatus\cite{TOTEM_JINST}, composed of three subdetectors (Roman Pots (RP), T1 and T2 telescopes), is placed symmetrically on both sides of Interaction Point (IP) 5, shared with the CMS experiment. All three subdetectors have trigger capability.
The Roman Pot stations, equipped with silicon detectors and placed at 147 and
220 m from the IP, detect elastically and diffractively scattered protons with a small scattering angle down to a few $\mu$rad. The T1 and T2 telescopes, placed at about 8 and 14 m from the IP respectively, detect charged particles produced in the polar angular range of a few mrad to $\sim\,$100 mrad. The T1 telescope (3.1 $<|\eta|<$ 4.7) consists of Cathode Strip Chambers, while the T2 telescope (5.3 $<|\eta|<$ 6.5) is made of triple-GEM (Gas Electron Multipliers) chambers \cite{SAULI}. The present analysis is based on measurements with the T2 detector that consists of 2 quarters with 10 semicircular chambers each, on both sides of the IP.
Each chamber provides two-dimensional information of the track position in an azimuthal coverage of \(192\,^{\circ}\) with a small overlap region along the vertical axis between chambers of two neighbouring quarters\cite{triple-GEM}. Every chamber has a double layered read-out board containing two columns of 256 concentric strips (\(400\,\mu m\) pitch, \(80\,\mu m\) width) for the measurement of the radial coordinate and a matrix of 1560 pads, each one covering $\Delta\eta \times \Delta\phi\ \simeq 0.06 \times 0.018$ rad, for the measurement of the azimuthal coordinate and for triggering. Radial and azimuthal coordinate resolutions are about 110$\,\mu$m and 1$\,^{\circ}$, respectively\cite{triple-GEM2}. The total material of 10 chambers amounts only to $\sim\,$0.05 X$_{0}$ \cite{EraldoPhd}. The read-out of all TOTEM detectors is based on the ``VFAT'' front-end ASIC, which provides as output a digital signal and trigger \cite{VFAT}.

\section{Detector simulation}\label{SectionTrackreco}
The TOTEM software\cite{CMSSW}, based on the CMS framework\cite{CMSSWcms}, embeds the necessary interfaces to the GEANT4\cite{GEANT4} simulation toolkit, the description of the material placed between the IP and T2 and the offline software used for the event reconstruction. The description of the T2 detector is implemented within this framework. 
The GEM signal digitisation has been parametrised using a dedicated model that, after proper tuning, reproduces well the measured cluster (a group of neighbouring strip or pad channels) size and reconstruction efficiency as a function of the ionisation energy released in the gas by the incident particle, diffusion coefficient of the fill gas, chamber gain and the VFAT thresholds\cite{EraldoPhd,MirkoTesi}. 
A special effort was devoted to understand and quantify secondary particles produced by the interaction of particles with the material in front of and around T2 and then seen in the detector. The simulation of the forward region, properly tuned with the data, showed that a large number of secondary particles is produced in the vacuum chamber walls in front of the detector, in the beam pipe conical section at \(|\eta| = 5.53\) and at the lower edge of the CMS Hadron Forward (HF) calorimeter \cite{PaoloTesi}.  

\section{Track reconstruction and alignment}
The track reconstruction is based on a Kalman Filter-like algorithm\cite{MirkoTesi} that is simplified due to the small amount of material traversed by the particle crossing the 10 GEM planes and to the low local magnetic field in the T2 region. The particle trajectory can, therefore, be successfully reconstructed using a straight line fit. The reconstructed tracks have at least 4 hits (pad clusters with or without an overlapping strip cluster), of which at least three have a pad/strip cluster overlap. A $\chi^{2}$-probability greater than $1\%$ is required for the straight line fit. 

\begin{figure}[htb]
\centering
\includegraphics[height=3.0in, width=0.91\linewidth]{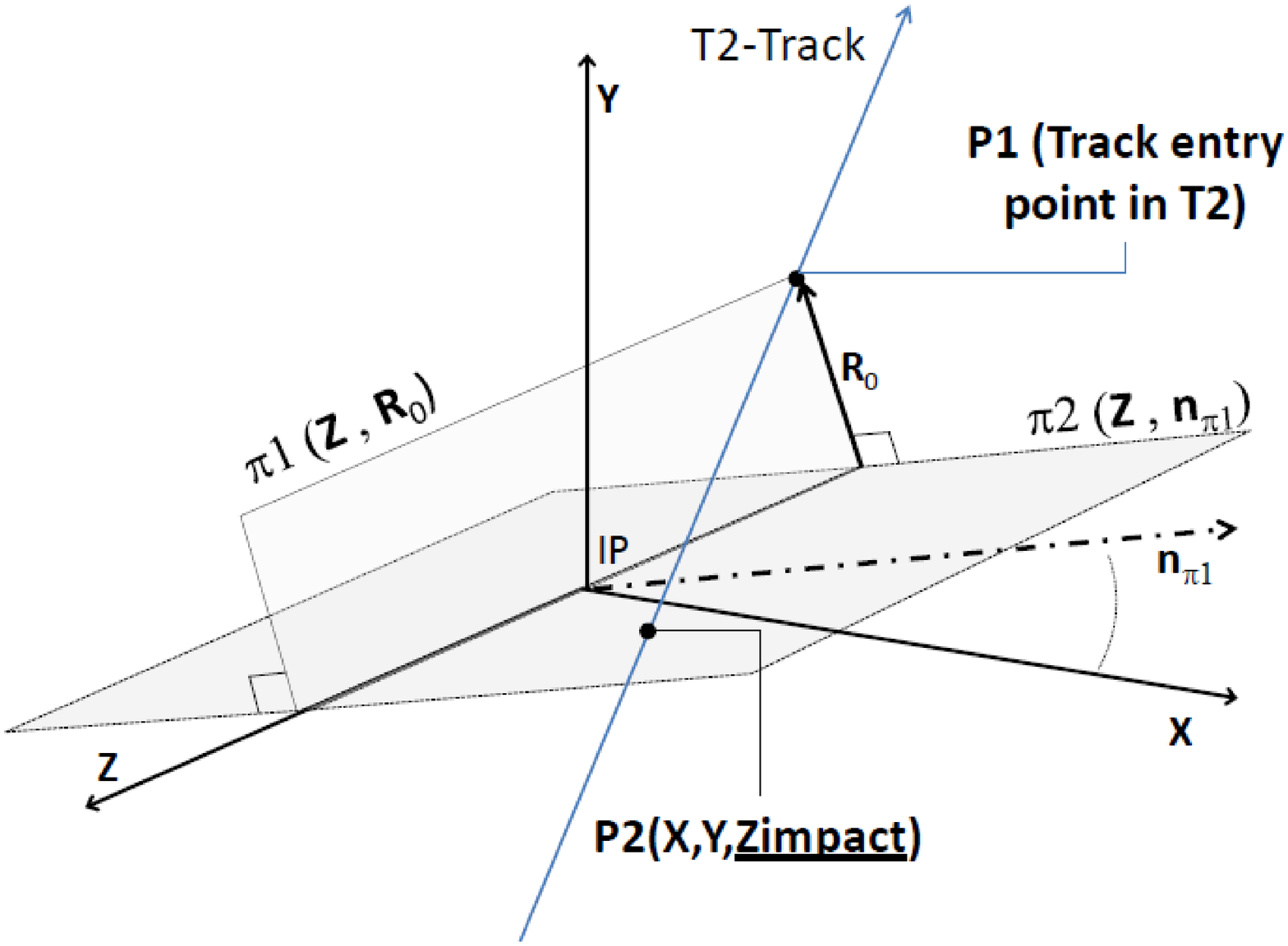}
\caption{Definition of the track ZImpact parameter. The description is discussed in the text.
}\label{MCPrimSecfitsBeforeAftera}
\end{figure}

Using a coordinate system with the origin located at the nominal collision point, the X axis pointing towards the centre of the LHC ring, the Y axis pointing upward (perpendicular to the LHC plane), and the Z axis along the counterclockwise beam direction, two geometrical track parameters are defined: Z$_{0}$ and ZImpact. Z$_{0}$ is the Z  value at the position of the minimum approach of the track to the Z axis; the ZImpact parameter (see fig. \ref{MCPrimSecfitsBeforeAftera}) is the Z coordinate of the intersection point between the track and a plane (``$\pi_{2}$'') containing the Z axis and orthogonal to the plane defined by the Z axis and the track entry point in T2 (``$\pi_{1}$'').
Due to the short lever arm of the T2 detector ($\sim\,$40 cm) compared to the distance to the IP, the ZImpact and Z$_{0}$ resolution is of the order of 1 m.

The relative position of the detector planes within a T2 quarter  (internal alignment) and the overall alignment of all the detector planes with respect to their nominal position (global alignment) have been investigated in detail to define possible misalignment biases of the track measurements\cite{MirkoTesi}. The most important internal alignment parameters which can be resolved within the T2 hit resolution are the shifts of the planes in the X and Y directions. Two different methods (iterative and MILLIPEDE \cite{Millipede}) were used to correct for such displacements. Both gave consistent results, with an uncertainty on the transverse position of the plane of about 30$\,\mu$m. The relative alignment between the two neighbouring quarters was obtained using tracks reconstructed in the overlap region.

The global alignment of the detector is of main importance for the present analysis. This was achieved by exploiting the symmetric distribution of the track parameters and the position of the ``shadow'' of the beam pipe, a circular shaped zone of the T2 planes characterised by a very low hit rate due to interactions of primary particles in the $|\eta|\,=\,$5.53 beam pipe cone in front of T2.

The combination of these methods gave, for each quarter, the X-Y shift with respect to the nominal position with a precision of $\sim\,$1 mm and the tilts in the XZ and YZ planes with a precision of $\sim\,$0.4 mrad. %Variation was  of 0.3 mrad, 

After the local and global alignment parameters had been measured with the data, the corresponding misalignments were introduced into the GEANT4 simulation and the same algorithm for the correction of the hit positions was applied to the reconstruction of both simulation and data. In this way one can also take into account, in the simulation, the non-uniform effect that misalignment has on the reconstructed hit position, which depends on its X-Y coordinate in the GEM plane.
%The effect of the central CMS magnetic field and the material interaction on the propagation of the particle is evaluated using the CMS software framework. Look at EnergyCutSystematic histo in /afs/cern.ch/exp/totem/scratch/berretti/tmp/Analysis/Ntuple_Analysis/Recorrection_Alignment_RootFiles

\section{Data and MC sample}\label{samples}
The sample used for the present analysis consists of 150,000 $pp$ collisions at $\sqrt{s}= 7$ TeV recorded in May 2011 during a low pile-up run at standard optics. Each beam was composed of six bunches with an average luminosity per colliding bunch pair of about 8$\,\times\,10^{27}\,$cm$^{-2}\,$s$^{-1}$ corresponding to an inelastic pile-up probability of  $\sim\,$3\%. The rate of beam gas interactions for such beam conditions is expected to be negligible.
The trigger required at least one trigger-road, defined as more than 3 ``superpads'' (3 radial and 5 azimuthal neighbouring pads) fired in the same r-$\phi$ sector of different planes of the same T2 quarter. This condition is satisfied if at least one charged particle traverses the T2 detector. The observed trigger rate  was about 3 kHz. The fraction of dead or noisy channels in this data sample has been measured to be 6\% for the pads and 9.5\% for the strips.
%run 5601: bunch crossing  trigger (T1 and T2 active). For standard T2 trigger I'm confident on the good data quality of the small bunches runs 5530 and 5525

With the requirement of at least one reconstructed  track in the T2 detector, the visible cross section seen by T2 has been estimated to be about 95\% of the total inelastic cross section. This is based on the comparison of the direct measurement of the T2 visible inelastic cross section (to be published) to the TOTEM inelastic cross section measurement deduced from the difference between  the total and elastic cross sections\cite{totalTOT}. The fractions of the total inelastic cross section visible to T2 obtained from Pythia 6.42\cite{Pythia6} and Pythia 8.108\cite{Pythia8} are in agreement with the above estimate. The T2-triggered sample contains more than 99\% of all non-diffractive events and all  single and double diffractive events having at least one diffractive mass larger than $\sim\,$3.4 GeV/c$^{2}$\cite{PaoloTesi}. 

The transverse momentum ($P_{T}$) acceptance for single charged particles going into T2 is limited by the magnetic field and multiple scattering effects. Simulation studies have shown that the charged particle tracks are reconstructed, within the analysis cuts utilised in this work, with a good efficiency for $P_{T}\geq\,$40 MeV/c, defining effectively the minimum $P_{T}$ acceptance. 
The fraction of charged particles with $P_{T}<\,$40 MeV/c produced in the T2 acceptance is predicted to be very small ($\sim\,$1\%). 
%Particles with lower $P_{T}$, corresponding to ($\sim\,$1\%), are reconstructed with an efficiency below 80\%.

\section{Analysis procedure}\label{datareduction}
\begin{widetext}
\begin{equation}
\label{s.long}
\frac{dN}{d\eta}\Bigg\arrowvert_{\eta=\eta_{0}}=\sum_{\mathrm{Trk} \in\, S}\,\,\dfrac{\,W(\eta_{0},\mathit{ZImpact})\,\sum_{j}B_{j}(\eta_{0})}{\epsilon\,(\eta_{0},m)\,\Delta\eta\,N_{\mathrm{Ev}}}\,G(\eta_0{})\,S_{p}(\eta_{0})\,\dfrac{2\pi}{\mathit{\Phi}}\,H\,P
\end{equation}
\end{widetext}

Our pseudorapidity density measurement refers to charged particles with a lifetime longer than $0.3\,\times\,10^{-10}$ s, and to the charged decay products of particles with shorter lifetime, which is consistent with the ATLAS\cite{ATLASref}, ALICE\cite{ALICEref} and CMS\cite{CMSref} definition of a primary charged particle. With this definition, decay products of the $K^{0}_{s}$ and $\Lambda$ hadrons are considered secondary particles, together with all of the charged particles generated by interactions with the material in front and around T2.
The $\eta$-value of a track is defined here as the average pseudorapidity of the T2 track hits, calculated from the angle that the hit has with respect to the beam at the IP. This definition has been adopted on the basis of detailed MC simulation studies to find the optimal definition of the true $\eta$ of a particle produced at the IP. The pseudorapidity density has been measured for each quarter independently, allowing an important consistency check among the four analysis results, as each quarter differs in its alignment and track reconstruction efficiency. 

\begin{figure}[htb]
\centering
\includegraphics[height=3.2in, width=0.94\linewidth]{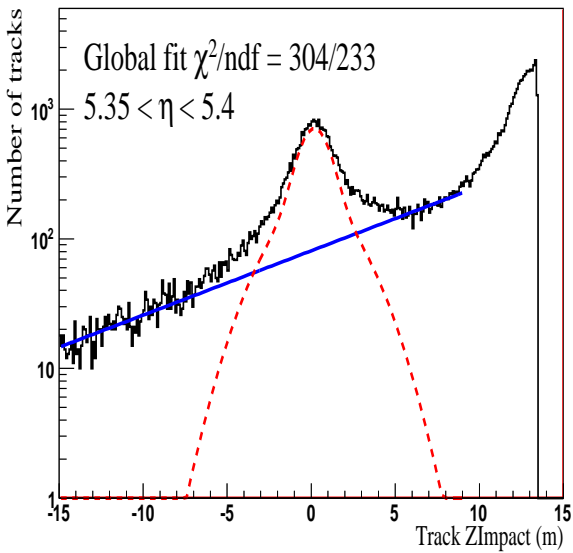} %H1LowEta2.eps H1LowEtaSH2.eps
\caption{ZImpact parameter distribution for the data tracks reconstructed in one T2 quarter in the $5.35<\eta<5.4$ range. The reported $\chi^2$/ndf refers to the global (double Gaussian + exponential) fit, performed in the range from -15 m up to 9 m. The blue-solid curve represents the exponential component due to secondaries, while the red-dashed curve is the double Gaussian component mainly related to primary tracks. The condition Z$_{0}\, \cdot$ sign($\eta$)\,$<13.5\,$m has been required, reducing the amount of secondary tracks by about 60\%.%Was 20 but is 60, far away from interaction region
}
\label{MCPrimSecfitsBeforeAfterb}
\end{figure}

Since about 80\% of the T2 reconstructed tracks are secondaries, it is important to have a procedure for the discrimination between them and primary charged particles. Based on detailed simulation studies, the most effective primary/secondary particle separation is achieved using the ZImpact track parameter\cite{MirkoTesi}.  This parameter is proven to be stable against misalignment errors and is well described by a double Gaussian function for the primary particles and by an exponental function for the secondaries.
The track ZImpact distribution, with the exponential and double Gaussian fit, is shown in fig. \ref{MCPrimSecfitsBeforeAfterb} for data tracks reconstructed in one T2 quarter in the  $5.35<\eta<5.4$ range.

The mean, required to be the same for both Gaussians, the standard deviation and the amplitude of the two Gaussians for primaries as well as the mean and the amplitude of the exponential for secondaries have been left free in the fit. Since the fit results have been found to be $\eta$-dependent, the fit, performed on data, has been repeated for each $\eta$-bin of the pseudorapidity distribution giving standard deviations (amplitudes) of both Gaussians that increases (decreases) with $\eta$. The relative abundance of secondary particles has been found to be smaller for higher $|\eta|$.

The primary tracks were selected using the ZImpact and Z$_{0}$ parameters: ZImpact was required to be in the range for which the area covered by the double Gaussian is 96\% of the total, while Z$_{0}$ was required to fulfil the condition Z$_{0}\, \cdot$ sign($\eta$)\,$<13.5\,$m.
The fraction of primary tracks, among the ones passing the above selection criteria, was calculated for each $\eta$-bin as a function of the ZImpact-value using the double Gaussian and exponential fits. This fraction, found to range from about 75\% (lower $|\eta|$ bins) to about 90\% (higher $|\eta|$ bins),  allows each data track to be weighted by the probability for the track to be primary, according to its $\eta$ and ZImpact-value. 

Each track has then been weighted for the primary track efficiency according to its $\eta$ and to the pad-cluster multiplicity in the corresponding quarter. This efficiency, evaluated by MC, is defined as the probability to successfully reconstruct a GEANT4 generated primary track that traverses the detector and yields the Z$_{0}$ and ZImpact parameters within the allowed region. The dependence of the efficiency from the pad-cluster multiplicity was included to make this correction independent from the tuning of the MC multiplicity. Once the combined Z$_{0}$ and ZImpact requirements were applied, an average primary track efficiency of $\sim\,$80\% was obtained.

A small contribution to the double Gaussian peak is given by the decay products of strange particles and by conversion of photons from $\pi^{0}$ decays in the material near T2. The overall non-primary contribution to the central peak, to be subtracted, has been estimated using different MC generators and found to range between 6\% and 13\% according to Pythia, Phojet\cite{Phojet} and Sherpa\cite{Sherpa}. For improving the description of the non-primary fraction of the tracks, the $K^{0}_{S}$ dN/d$\eta$ and $\gamma$ dN/dE in the MC were normalised in order to reproduce, in their acceptance region, the measurements by CMS\cite{cmsk0} and LHCf\cite{lhcfgamma}.

MC studies also provided the fraction of primary charged particles which do not arrive in the T2 detector. The associated correction factor, in average $\sim\,$1.04, has been calculated for each $\eta$ bin in events with at least a charged particle in the $5.3<|\eta|<6.5$ range, by considering the number of primary GEANT4 tracks crossing the detector and the corresponding number of primary charged particles generated at the IP with P$_T>\,$40 MeV/c.

A bin migration correction accounting for all smearing effects on the reconstructed track $\eta$ was also derived. The track distribution was then normalised to the full acceptance in azimuthal angle. The pile-up probability, determined from data, was finally subtracted by means of an overall correction factor of 0.97.

Events characterised by a high hit multiplicity, typically due to showers generated in interactions with the material, were not included in the analysis. These events, where track reconstruction capability is limited, constitute about 11\% of the data sample, and have an average pad cluster multiplicity per plane larger than 70. The effect of not considering these events has been evaluated in a MC study, giving an overall correction factor on the measurement of about 1.02. %13-10-10-10

Eq. (\ref{s.long}) was used for the dN$_{\textnormal{ch}}$/d$\eta$ determination: $\eta_{0}$ is the $\eta$-value of the bin centre, S is the sample of tracks with $\eta_{0}-\Delta\eta/2<\eta<\eta_{0}+\Delta\eta/2$ satisfying the selection criteria above, $\Delta\eta=0.05$ is the bin width, $N_\textrm{Ev}$ is the number of events in the data sample, $W$ is the probability for a track to be primary, $\epsilon$ is the primary track efficiency (where $m$ indicates the event pad-cluster multiplicity), $B_j$ is the bin migration correction associated with the $j$th bin, $G$ is the correction factor for primary particles not reaching T2, $S_{p}$ is the correction factor for the non-primary contribution to the double Gaussian peak, $\mathit{\Phi}/2\pi$ is the azimuthal acceptance, $H$ is the correction factor taking into account the effect of the exclusion of the events with high secondary multiplicity and $P$ is the pile-up correction factor.

\section{Systematic uncertainties}\label{SystError}
The systematic uncertainty associated with the primary track efficiency has been evaluated in studies where tracks were reconstructed using a set of  5 detector planes (out of the total of 10) in a single T2 quarter. The track reconstruction efficiency was determined using the other set of 5 detector planes in the same quarter. The associated systematic uncertainty, estimated with this procedure, was defined as the difference between the result obtained using the above data driven method, and the MC analysis using the same definitions. This uncertainty, computed as a function of the pad-cluster multiplicity and of the track $\eta$, has been found to give a relative contribution from 1 to 7 \%.

In order to evaluate the systematics due to the alignment corrections, the global alignment parameters have been varied around the optimal values within their resolution. The data have been again reconstructed and analysed for the different misalignment configurations. The corresponding variation in the dN$_{\textnormal{ch}}$/d$\eta$ result defined the systematics due to the alignment. This uncertainty, which is $\eta$ and quarter dependent, has been found to be in the range of 3-4\%.

The systematic uncertainty associated with the fraction of the non-primary contribution to the central peak, $S_{p}$, has been evaluated by considering the maximum variation obtained from several MC generators. It has been found to be $\eta$-dependent, ranging from 1\% to 3\%.

The systematic uncertainty on the $W$ function, needed for the primary to secondary separation, has been estimated to be in the range of 2-3\% considering the uncertainty on the fitting parameters.

The primary track efficiency variation due to magnetic field effects and to the uncertainty on the energy spectrum resulted in an error of about 2\%. This contribution has been evaluated in the simulation by switching on and off the magnetic field and by varying in a reasonable range the energy spectrum\footnote{The CMS experiment reported a discrepancy, that increases with $|\eta|$, between data and Pythia inelastic event simulation \cite{CMSEnergy}. Pythia underestimates the energy flow in the CMS HF calorimeter, measured in the pseudorapidity range $3.15<|\eta|<4.9$. The effect of this discrepancy on the reported analysis has been taken into account using a dedicated simulation, where the input energy spectrum of the particles has been increased according to the extrapolated discrepancy expected in the $5.3<|\eta|<6.5$ region.
}.

The variation of the $G$ function from the MC generator has been  estimated to be around 2\%, while the $B_{j}$ functions have been found to have a negligible variation.
The effect of the track quality criterion requirement, $\chi^{2}$-probability$\,>1\%$, has been estimated to be around 1\% by evaluating the data/MC discrepancy observed with and without using this requirement. The effect of the trigger bias in our measurement has been evaluated by comparing the data selected with a pure bunch crossing trigger with the sample triggered with T2 and has been found to be around 1\%. The pile-up probability systematic uncertainty has been estimated to be  1\%. The uncertainty on the correction accounting for the exclusion of events with high multiplicity of secondary particles has been found to be about 1\% from the difference between Pythia8 and Sherpa MC predictions.

Table \ref{onebinsysttab} shows the uncertainties of the bin centred at $\eta_{0}=6.025$, for one of the T2 quarters. The double dashed line separates the quarter dependent contributions (top) from the ones in common for all the quarters (bottom).
\begin{table}[!ht]
\caption{Summary of the relative uncertainties in the bin centred at $\eta_{0}=6.025$ in one of the T2 quarters. The first two contributions are quarter dependent.}
\label{onebinsysttab}
\begin{center}
\begin{tabular}{lc}
   \hline
\multicolumn{2}{c}{$\eta_{0}=6.025$ dN$_{\textnormal{ch}}$/d$\eta$ error summary (one quarter)} \\
   \hline
1. Primary track efficiency  &  4\%    \\ 
2. Global alignment  &  3\%    \\
\hdashline\hdashline
3. Non-primaries in the central peak &  2\% \\ 
4. Primary to secondary separation & 2\% \\
5. B-field and energy spectrum & 2\% \\
6. Primaries not arriving in T2 & 2\% \\
7. Track quality criterion & 1\% \\
8. Trigger bias & 1\% \\ 
9. Pile-up probability & 1\% \\
10. Events with high secondary multiplicity & 1\% \\
11. Statistical & 0.7\% \\
\hline

Total (single quarter measurement)& 10\% \\
\hline
\end{tabular}
\end{center}
\end{table}
The total systematic uncertainty has been computed by first linearly adding the global alignment and track efficiency systematics to take into account misalignment effects on the primary track efficiency estimation, then this result has been added in quadrature to the uncertainty contributions from 4) to 10) of table \ref{onebinsysttab} and finally the uncertainty associated with the non-primary contribution to the central peak has been added linearly. To obtain the total uncertainty of the single quarter measurement, the statistical error is then added in quadrature.

\section{Results}\label{sectionresults}
The dN$_{\textnormal{ch}}$/d$\eta$ measurements obtained for the different T2 quarters are compatible within the quarter-dependent systematic uncertainties.
For each $\eta$-bin, the measurements for the four quarters have been combined with a weighted average using only the quarter-dependent uncertainties. 

A conservative approach has been adopted for the combination of the quarter-dependent systematic uncertainties: an error propagation on the weighted averages has been applied, considering the measurements completely and positively correlated. The resulting error has then been combined with the systematic contributions that are common to all quarters and with the statistical one, as in the case of the single quarter measurement.

\begin{figure}[!ht]
\centering
\includegraphics[height=3.0in,width=1.0\linewidth]{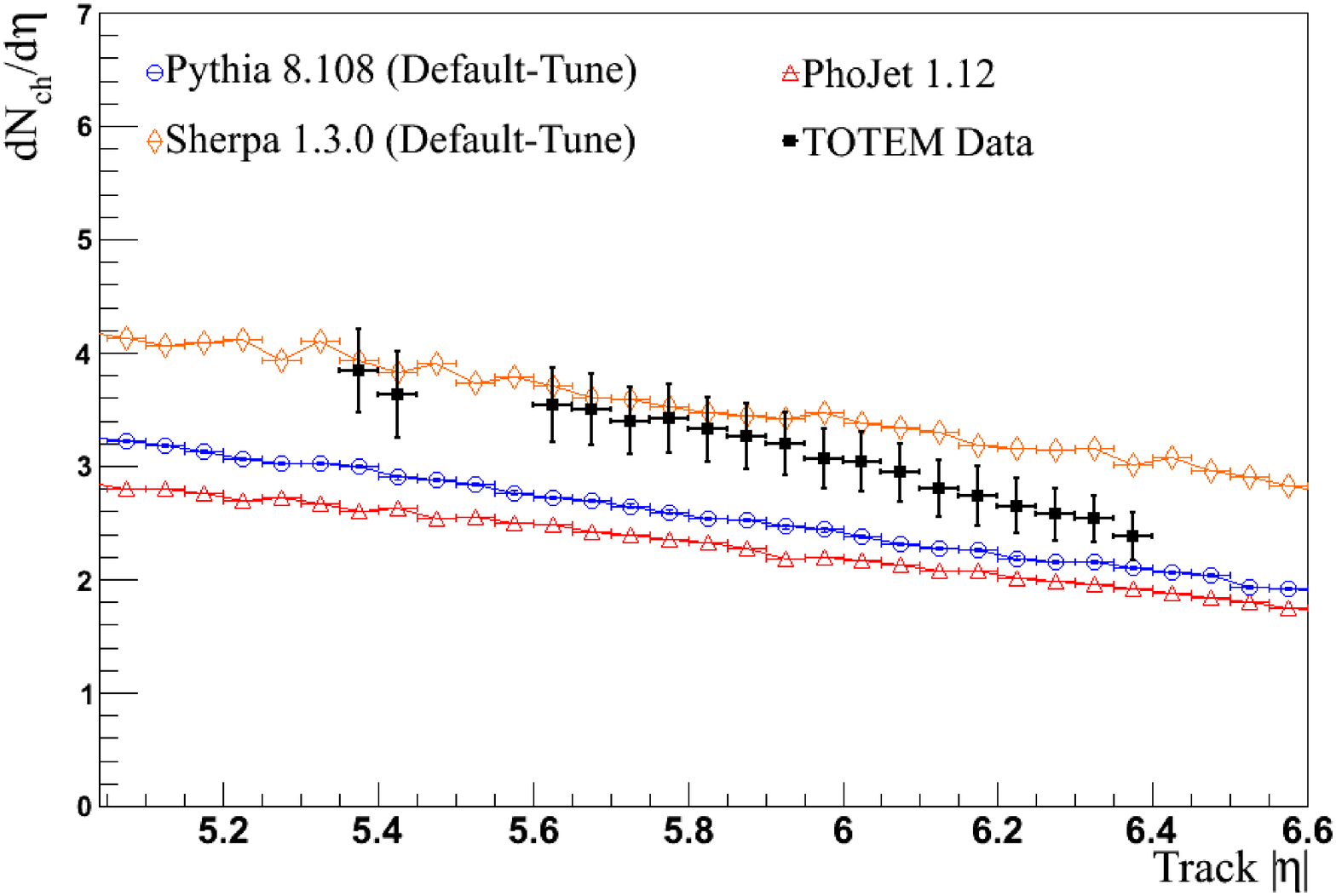}
\caption{Charged particle pseudorapidity density distribution. The experimental points (black squares) represent the average of the four T2 quarters, with the error bars including both statistical and systematic error. Red triangles, blue circles and orange diamonds show, respectively, the Phojet, Pythia8 and Sherpa predictions for charged particles with $P_{T}>\,$40 MeV/c in events where at least one charged particle is generated in the $5.3<|\eta|<6.5$ range.}
\label{dndetaPlot}\vspace{2em}
\end{figure}

The pseudorapidity density measurement is shown as black squares in fig. \ref{dndetaPlot} where the error bars represent the total uncertainty including the statistical error. The pseudorapidity bins at the edges of the T2 detector, where large corrections due to the reduced geometrical acceptance are needed, are not reported here, nor are the three bins in the $5.425\leq |\eta|\leq 5.625$ range, where large corrections should be applied because of the interaction of the primary particles with the beam pipe. Fig. \ref{dndetaPlot} also shows the comparison of the data with some MC expectations. Phojet 1.12 (red triangles) estimates a $\sim\,$30\% ($\sim\,$20\%) lower dN$_{\textnormal{ch}}$/d$\eta$ than measured at $|\eta|= 5.3$ (6.4).
Both Pythia 8.108 with default tune (blue circles) and Pythia 6.42 D6T estimate a  $\sim\,$20\% ($\sim\,$12\%) lower dN$_{\textnormal{ch}}$/d$\eta$ than measured at $|\eta|=5.3$ (6.4). Sherpa 1.3.0 with default tune (orange diamonds) is in agreement with the data for $|\eta|<5.9$, while it estimates a higher dN$_{\textnormal{ch}}$/d$\eta$ by $\sim\,$25\% at $|\eta|=6.4$. Similar trends in the comparison of the data with MC predictions have also been found by the ATLAS\cite{ATLASref}, ALICE\cite{ALICEref}, CMS\cite{CMSref} and LHCb\cite{LHCBref} experiments in their own $\eta$ acceptance. When comparing with cosmic ray MC generators, SYBILL\cite{syb} predicts a 4-16\% lower dN$_{\textnormal{ch}}$/d$\eta$, while QGSJET01\cite{qsgj1}, QGSJETII\cite{qsgj2} and EPOS\cite{epos} predict a 18-30\% higher dN$_{\textnormal{ch}}$/d$\eta$.

The dN$_{\textnormal{ch}}$/d$\eta$ measurement is also reported in table \ref{tabdndeta} for each $\eta$-bin with the corresponding systematic and statistical error.
\begin{table}[!ht]
\caption{
TOTEM dN$_{\textnormal{ch}}$/d$\eta$ measurement for inelastic pp events at $\sqrt{s}=$7 TeV. The reported values represent the average for the four T2 quarters, with the corresponding systematic (syst) and statistical (stat) error. $\eta_{0}$ represents the central pseudorapidity value in each $\eta$ bin.}
\label{tabdndeta}
\begin{center}
\begin{tabular}{cccc}
   \hline
$\eta_{0}$ $\,\,$ & dN$_{\textnormal{ch}}$/d$\eta$ & syst & stat  \\
    \hline
5.375  &  3.84  &  0.37  &   0.01  \\
5.425  &  3.64  &  0.38  &   0.01  \\
5.625  &  3.54  &  0.33  &   0.01  \\
5.675  &  3.50  &  0.32  &   0.01  \\
5.725  &  3.40  &  0.30  &   0.01  \\
5.775  &  3.42  &  0.31  &   0.01  \\
5.825  &  3.32  &  0.29  &   0.01  \\
5.875  &  3.27  &  0.29  &   0.01  \\
5.925  &  3.20  &  0.28  &   0.01  \\
5.975  &  3.07  &  0.27  &   0.01  \\
6.025  &  3.04  &  0.26  &   0.01  \\
6.075  &  2.94  &  0.26  &   0.01  \\
6.125  &  2.80  &  0.25  &   0.01  \\
6.175  &  2.74  &  0.26  &   0.01  \\
6.225  &  2.65  &  0.24  &   0.01  \\
6.275  &  2.58  &  0.23  &   0.01  \\
6.325  &  2.53  &  0.21  &   0.01  \\
6.375  &  2.38  &  0.21  &   0.01 
\end{tabular}
\end{center}
%\vspace{1em}
%\vspace{-2em}
\end{table}

%\flusbottom

\section{Conclusions}
The TOTEM experiment has measured the charged particle pseudorapidity distribution in $pp$ 
collisions at $\sqrt{s} =$ 7 TeV for $5.3<|\eta|<6.4$ in events with at least one reconstructed track in this range. 
This extends the  measurements performed by the other LHC experiments to this previously unexplored forward $\eta$ range.
The measurement refers to charged particles with $P_{T}>\,$ 40 MeV/c and with a mean lifetime $\tau > 0.3\times 10^{-10}$ s, directly produced in $pp$ interactions or in subsequent decays of particles having a shorter lifetime.
A preliminary measurement of the T2 visible inelastic cross section confirms that
about 95\% of the inelastic $pp$ events have been considered in the present study.
This comprises more than 99\% of non-diffractive processes and the single and double diffractive processes with diffractive masses above $\sim\,$3.4 GeV/$c^{2}$. The pseudorapidity density has been found to decrease with increasing $|\eta|$, from 3.84 $\pm$ 0.01(stat) $\pm$ 0.37(syst) at $|\eta| = 5.375$  to 2.38 $\pm$ 0.01(stat) $\pm$ 0.21(syst) at $|\eta| = 6.375$. Several MC generators have been compared to data; none of them has been found to fully describe the measurement.

\begin{center}
$\ast\ast\ast$
\end{center}
%\acknowledgments
We thank {\scshape M. Ferro-Luzzi} and the LHC machine coordinators for scheduling and providing us the dedicated TOTEM runs. We are very grateful to the CMS collaboration for providing us the software framework where all the toolkits used for the analysis reported here have been developed. We express gratitude to {\scshape R. Ulrich}  and  {\scshape C. Baus}  for providing us the cosmic ray MC predictions.

This work was supported by the institutions listed on the front page and partially also by NSF (US), the Magnus Ehrnrooth foundation (Finland), the Waldemar von Frenckell foundation (Finland),  the Academy of Finland, the OTKA grants NK 73143 and NK 101438 (Hungary).

\bibliographystyle{eplbib}

\bibliography{ArxivDNdeta}

\end{document}